\documentclass[12pt,a4paper]{article}
\usepackage{siunitx}

\usepackage[british]{babel}
\usepackage[utf8]{inputenc}
\usepackage{newunicodechar}
\newunicodechar{−}{\textminus}
\newunicodechar{⁴}{\textsuperscript{4}}
\newunicodechar{³}{\textsuperscript{3}}
\usepackage[a4paper,top=2cm,bottom=2cm,left=2.5cm,right=2.5cm,marginparwidth=1.75cm]{geometry}

\usepackage{xcolor}

\usepackage{setspace} 
\usepackage[normalem]{ulem}

\usepackage{amsmath}
\usepackage{graphicx}
\usepackage[colorlinks=true, allcolors=blue]{hyperref}
\usepackage{hyperref}
\usepackage{orcidlink}
\usepackage[title]{appendix}
\usepackage{mathrsfs}
\usepackage{amsfonts}
\usepackage{booktabs} 
\usepackage{caption}  
\usepackage{threeparttable} 
\usepackage{algorithm}
\usepackage{algorithmicx}
\usepackage{algpseudocode}
\usepackage{listings}
\usepackage{enumitem}
\usepackage{chngcntr}
\usepackage{booktabs}
\usepackage{lipsum}
\usepackage{subcaption}
\usepackage{authblk}
\usepackage[T1]{fontenc}    
\usepackage{csquotes}       
\usepackage{diagbox}
\usepackage{multirow}

\usepackage{cite}

\usepackage{setspace}
\usepackage{titlesec}
\titleformat{\section} 
  {\normalfont\Large\bfseries}{\thesection.}{1em}{}
  
\usepackage{float}   
\usepackage{caption} 


\title{\textbf{ \textit{Ab initio} Study of Substitutional Defects in Li$_{3}$OCl Solid Electrolyte for Li-ion Batteries}}

\author[2,3]{Carson D. Ziemke}
\author[3]{Naveed Naemi}
\author[1,2,5*]{Ha M. Nguyen}
\author[3] {Anthony Dorhauer}
\author[4]{Carlos Garcia}
\author[2]{Narendirakumar Narayanan}
\author[1,4]{Yangchuan Xing}
\author[2]{John Gahl}
\author[1,2,3]{Thomas W. Heitmann}
\author[1,3,*]{Carlos Wexler}
\affil[1]{\small Materials Science and Engineering Institute, University of Missouri, Columbia, MO 65201, USA}
\affil[2]{ University of Missouri Research Reactor, University of Missouri, Columbia, MO 65203, USA}
\affil[3]{Department of Physics and Astronomy, University of Missouri, Columbia, MO 65201, USA}
\affil[4]{Department of Chemical and Biomedical Engineering, University of Missouri, Columbia, MO 65201, USA}
\affil[5]{Center for Materials Innovation and Technology, VinUniversity, Hanoi, Vietnam}

\affil[*]{Corresponding authors: \texttt{wexlerc@missouri.edu,  hn4gq@missouri.edu}}

\begin{document}


{
\maketitle

\begin{abstract}
Improving ion transport in solid electrolytes and cathode coatings remains a key challenge for all-solid-state Li-ion batteries because their room-temperature ionic conductivity is still substantially lower than that of liquid electrolytes. In our previous combined experimental and theoretical study, we showed that thermal neutron irradiation enables defect engineering in LiBO$_2$ through the transmutation of $^6$Li and $^{10}$B, generating lattice vacancies that enhance ionic conductivity. Here, we examine whether this approach can be extended to Li$_3$OCl, a representative antiperovskite solid electrolyte. Using density functional theory, we investigate substitutional defects at Li sites involving B, He, and H, associated with B doping and the neutron-capture reactions $^{6}\mathrm{Li}+n\rightarrow\,^{3}\mathrm{H}+\alpha$ and $^{10}\mathrm{B}+n\rightarrow\,^{7}\mathrm{Li}+\alpha+\gamma$. We evaluate defect formation energetics, the resulting structural distortions, and compare these substitutional defects with other mono-, di-, and trication substitutions at Li sites. Our results show that substitutional defects associated with neutron irradiation provide a feasible route to tune the defect chemistry of antiperovskite solid electrolytes and support neutron-driven defect engineering as a strategy for developing advanced materials for high-performance all-solid-state Li-ion batteries.

\begin{figure*}[!ht]
\centering
\includegraphics[width=0.8\linewidth]{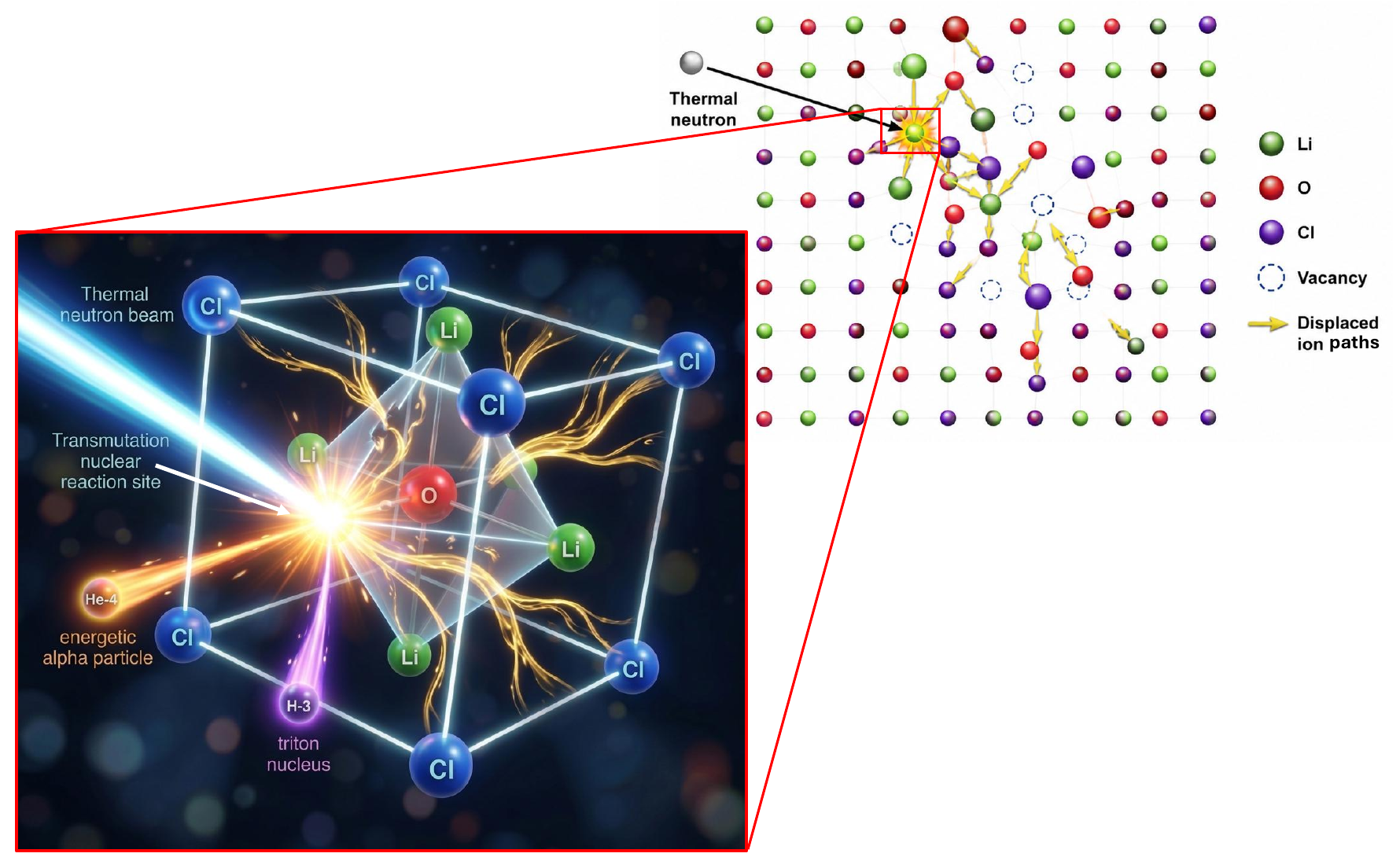}

\end{figure*}
\end{abstract}


\section{Introduction \label{Intro}}

\noindent
Lithium-ion batteries (LIBs) have emerged as a key pillar of modern energy storage systems, powering applications ranging from portable electronics to electric vehicles and grid-scale energy storage systems \cite{H_Wang_2023,JA_Dawson_2021,N_Nitta_2015,A_Dutra_2023,Z_Chen_2025}. 
However, conventional LIBs based on liquid electrolytes suffer from flammability, leakage, limited electrochemical stability, and safety concerns under thermal or mechanical abuse conditions \cite{Finegan2018,Schipper2016,Chen_2021}.
These limitations have driven extensive interest in all-solid-state lithium-ion batteries (ASSLIBs), where solid-state electrolytes (SSEs) replace liquid electrolytes to improve safety, thermal stability, and energy density \cite{Ramasubramanian2024}.
Among the many candidate SSEs, lithium-rich antiperovskites are promising because of their high lithium content, simple crystal structures, and potential for fast lithium-ion transport \cite{Y_Zhao_2012,Ghosh2022}.

\begin{figure}[!ht]
\centering
\includegraphics[width=0.6\linewidth]{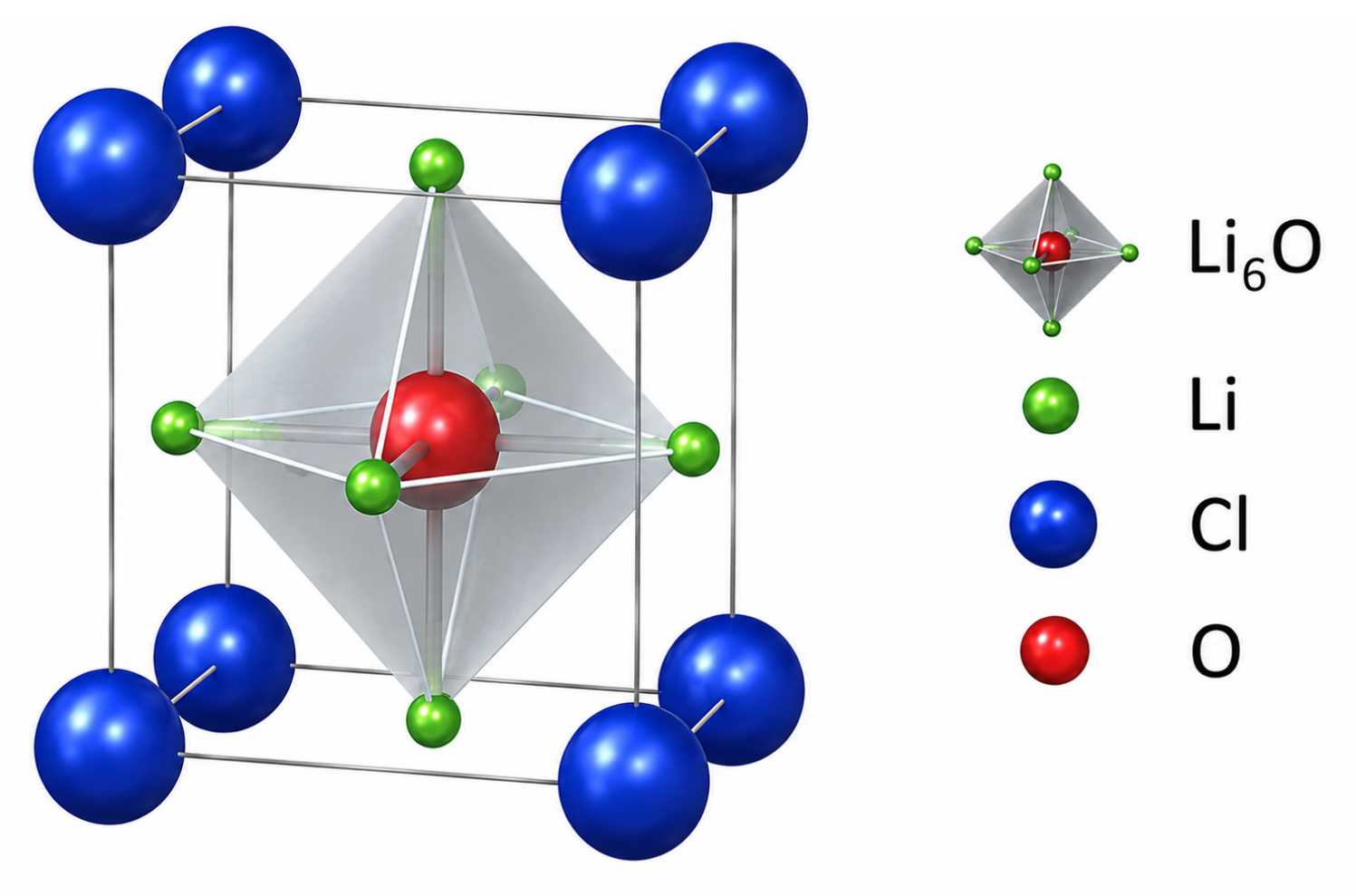}
\caption{\label{Figure_Li3OCl_crystal} Cubic crystal structure of lithium oxychloride Li$_{3}$OCl with space group  $Pm\bar{3}m$ and lattice constant 3.88 Å.}
\end{figure}

Lithium oxychloride (Li$_3$OCl) is one of the most extensively investigated lithium-rich antiperovskite SSEs \cite{Y_Zhang_2026,Zheng2025,Zheng2021,Dawson2012}.
It crystallizes in a cubic antiperovskite structure (space group $Pm\bar{3}m$), with O at the body center, Cl at the cube corners, and Li at the face-centered sites (Figure \ref{Figure_Li3OCl_crystal}). This highly symmetric framework supports a lithium sublattice favorable for ion migration.

In addition to its structural simplicity, Li$_3$OCl exhibits attractive properties for solid-state battery applications, including a high lithium concentration and a relatively wide electrochemical stability window \cite{Zhang2023}.  Li$_3$OCl has therefore been used as a model system for investigating ion transport, defect chemistry, and lattice dynamics in antiperovskite solid electrolytes \cite{Lu2015}.

Previous experimental and theoretical studies have shown that ion transport in Li$_3$OCl is strongly controlled by point defects and local lattice distortions. Emly \textit{et al.} \cite{A_Emly_2013} demonstrated that native lithium vacancies and interstitials play a key role in lithium-ion conduction, with defect formation energetics being strongly dependent on synthesis conditions. 
Zhang \textit{et al.} \cite{Y_Zhang_2013} further examined defect thermodynamics and aliovalent doping in Li$_3$OCl, highlighting defect engineering as an effective route for tuning ionic conductivity and phase stability.
Although substitutional doping has been widely studied in solid electrolytes, most work has focused on equilibrium dopants introduced during synthesis, whereas substitutional defects generated under irradiation or other non-equilibrium conditions remain much less explored.

We have recently introduced neutron irradiation as a potential tool for defect engineering in lithium-containing ionic solids. In our previous combined experimental and theoretical studies of LiBO$_2$, we showed that thermal-neutron irradiation can induce transmutation-driven lattice defects through neutron capture by $^{6}$Li and $^{10}$B, producing lattice vacancies and enhancing ionic conductivity via defect-mediated lithium-ion transport \cite{CD_Ziemke_2025,HaMNguyen2025b,HaMNguyen2026}. 
These results raise the question of whether neutron-assisted defect engineering can be extended to other solid-state lithium-ion conductors. In particular, neutron transmutation may generate unusual substitutional species and local defect environments that are not readily accessible through conventional synthesis.  Understanding the stability and structural effects of such defects is therefore important for assessing the potential of neutron-driven defect engineering in solid-state battery materials.

In this work, we use first-principles density functional theory (DFT) to investigate substitutional defects in Li$_3$OCl associated with neutron-transmutation-related species. Specifically, we consider substitution at Li sites by B, He, and H, corresponding to defect species associated with neutron capture by $^{10}$B and $^{6}$Li. Natural lithium contains about 7.5\% $^{6}$Li, which undergoes the reaction $^{6}\mathrm{Li} + n \rightarrow \, ^3\mathrm{H} + \alpha$, while boron contains approximately 20\% $^{10}$B, which undergoes $^{10}\mathrm{B} + n \rightarrow \, ^7\mathrm{Li} + \alpha + \gamma$.
We evaluate the formation energetics and local structural distortions of these neutron-induced substitutional defects and compare them with those of conventional mono-, di-, and trication substitutional dopants at Li sites. This study clarifies how neutron-induced substitutional defects affect the defect chemistry and structural stability of lithium-rich antiperovskite solid electrolytes, and it provides a basis for assessing neutron irradiation as a route for defect engineering in solid-state battery materials.

\section{Computational Methods\label{Methods}}

First-principles density functional theory (DFT) calculations were performed using the Quantum ESPRESSO (QE) package \cite{P_Giannozzi_2017,P_Giannozzi_2019} to investigate substitutional lattice defects in Li$_3$OCl induced by neutron-transmutation-related species as well as conventional mono-, di-, and trication dopants. The calculations focused on evaluating relative defect energetics, local structural distortions, lattice and volume modifications, and symmetry breaking associated with substitutional defects at Li lattice sites.

The exchange–correlation interactions were treated within the Perdew--Burke--Ernzerhof (PBE) generalized-gradient approximation (GGA) \cite{JP_Perdew_1996}. Projector augmented-wave (PAW) pseudopotentials from the PSLibrary v1.0.0 \cite{PSLibrary} were employed for all atomic species, providing an efficient and reliable description of valence electrons in lithium-containing ionic solids.

The pristine Li$_3$OCl crystal structure was obtained from the Materials Project database \cite{MaterialsProject}. All calculations were performed using a $1\times1\times1$ periodic supercell containing three lithium atoms at (0,0,0.5), (0,0.5,0), and (0.5,0,0), one oxygen atom at (0.5,0.5,0.5), and one chlorine atom at (0,0,0) (see Figure \ref{Figure_Li3OCl_crystal}). The Li substitutional site is at (0.5,0.5,0).

The plane-wave cutoff energy was set to 100 Ry and the charge-density cutoff to 400 Ry. A $5\times5\times5$ Monkhorst--Pack $k$-point mesh was used. Convergence tests ensured total-energy accuracy within ~1 meV per supercell. Structural relaxations were performed using the BFGS algorithm \cite{R_Fletcher_1987} until forces were below $10^{-5}$ Ry/Bohr and total-energy changes below $10^{-5}$ Ry.

The pristine structure was first optimized using variable-cell relaxation (\texttt{vc-relax}) in QE. Substitutional defects were then generated by replacing one Li atom with a monocation (H, Na, K), a dication (Ba, Ca, He, Mg), or a trication (Al, B, Gd) species, denoted as A. Again, elements such as H, He, and B were particularly selected to respectively represent such specific neutron-transmutation-related isotopic species as alpha particles ($^{4}\text{He}$), tritons ($^{3}\text{H}$), and $^{10}$B, which arise from neutron-capture reactions involving $^{6}$Li and $^{10}$B. Each defective structure was fully relaxed using the same \texttt{vc-relax} procedure.

Defect formation energies were evaluated using

\begin{equation}
\label{FormationEnergyEquation}
E_{\mathrm{f}} =
E_{\mathrm{tot}}^{\mathrm{defect}}
-
E_{\mathrm{tot}}^{\mathrm{perfect}}
+
E_{\mathrm{Li}}
-
E_{\mathrm{A}},
\end{equation}

\noindent
where $E_{\mathrm{tot}}^{\mathrm{defect}}$ and $E_{\mathrm{tot}}^{\mathrm{perfect}}$ are the total energies of defective and pristine supercells, and $E_{\mathrm{Li}}$ and $E_{\mathrm{A}}$ are reference energies of Li and substitutional species A in their elemental states.  The H$_2$ molecule was used for H reference, while bulk phases were used for metals. A Gaussian smearing of 0.01 Ry was applied for metallic systems to aid convergence.

The $1\times1\times1$ supercell enables efficient and systematic screening of a wide range of substitutional defects while maintaining a consistent defect concentration and symmetry-breaking environment. Such reduced-supercell approaches are commonly used in comparative defect studies and descriptor-based materials screening \cite{Islam_2011}. This setup allows reliable comparison of relative trends in defect energetics, local structural distortions, and symmetry changes across chemically diverse dopants while remaining computationally tractable for large-scale screening relevant to neutron-induced defect engineering \cite{Deck_Hu_2023,CD_Ziemke_2025,HaMNguyen2025b}. For example, as reported by Islam \textit{et al.} \cite{Islam_2011} and discussed in our previous work \cite{CD_Ziemke_2025}, the calculated lithium vacancy formation energies in monoclinic LiBO$_2$ are 6.87 eV, 6.91 eV, and 6.98 eV for the 1$\times$1$\times$1, 2$\times$2$\times$2, and 3$\times$3$\times$3 supercells, respectively. Relative to the 1$\times$1$\times$1 supercell, the use of 2$\times$2$\times$2 and 3$\times$3$\times$3 supercells improves the calculated formation energy by approximately 0.6\% and 1.6\%, respectively. However, these modest gains in accuracy come at the expense of an exponential increase in computational cost as the supercell size grows. This tradeoff between computational accuracy and computational cost is illustrated more clearly in Figures \ref{Figure_DefectiveLi3OCl_formationLivacancy} and \ref{Figure_DefectiveLi3OCl_formationLivacancyComputationaltime}, using the lithium vacancy formation energy in Li$_3$OCl as an example. As shown in Figure \ref{Figure_DefectiveLi3OCl_formationLivacancy}, the DFT-calculated values (solid circles) of the formation energy ($E_{\text{f}}$) of lithium vacancy in Li$_3$OCl for the 1$\times$1$\times$1, 2$\times$2$\times$2, and 3$\times$3$\times$3 supercell is 14.714 eV, 14.550 eV, and 14.496 eV, respectively.  The dependence of $E_{\text{f}}$ on the reciprocal of the linear dimension of the supercells ($L$), as shown in Figure \ref{Figure_DefectiveLi3OCl_formationLivacancy} can be fit well ($R^{2}=0.997$) to a theoretical model (solid line) using a first-order approxmiation of a linear function of the form \cite{Mikael2018,Makov_Payne_1995}

\begin{equation}
\label{eqn2}
    E_{\text{f}} (L) = E^{({0})} +\frac{E^{({1})}}{L} + \mathcal{O}[\frac{1}{L^3}], 
\end{equation}

\begin{figure}[!ht]
\centering
\includegraphics[width=1.0\linewidth]{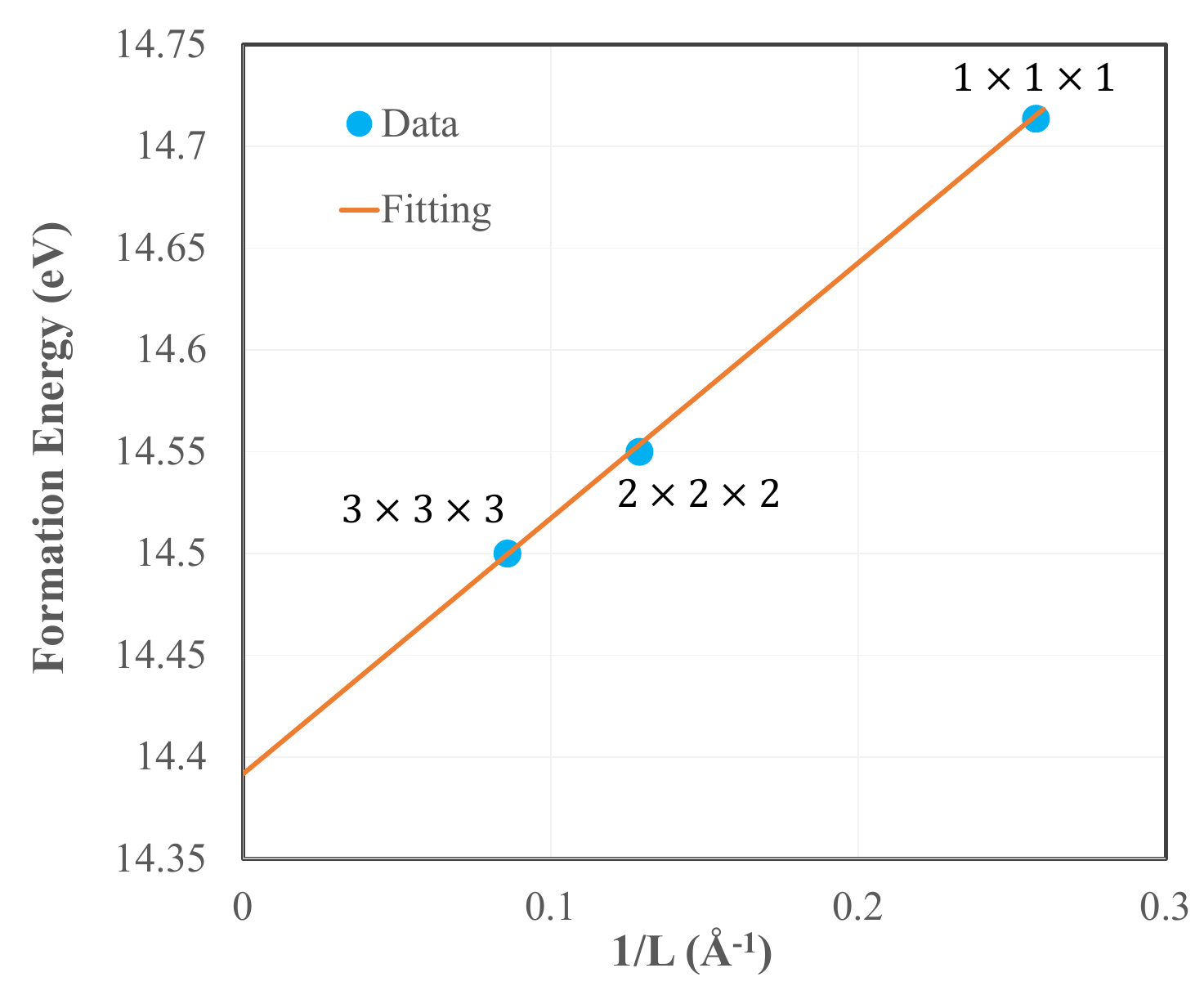}
\caption{\label{Figure_DefectiveLi3OCl_formationLivacancy} Dependence of the formation energy of lithium vacancy in Li$_{3}$OCl on the reciprocal of the linear size of calculation supercell $1/L$ (Å$^{-1}$). Each supercell is labeled next to its corresponding data point.}
\end{figure}

\begin{figure}[!ht]
\centering
\includegraphics[width=1.0\linewidth]{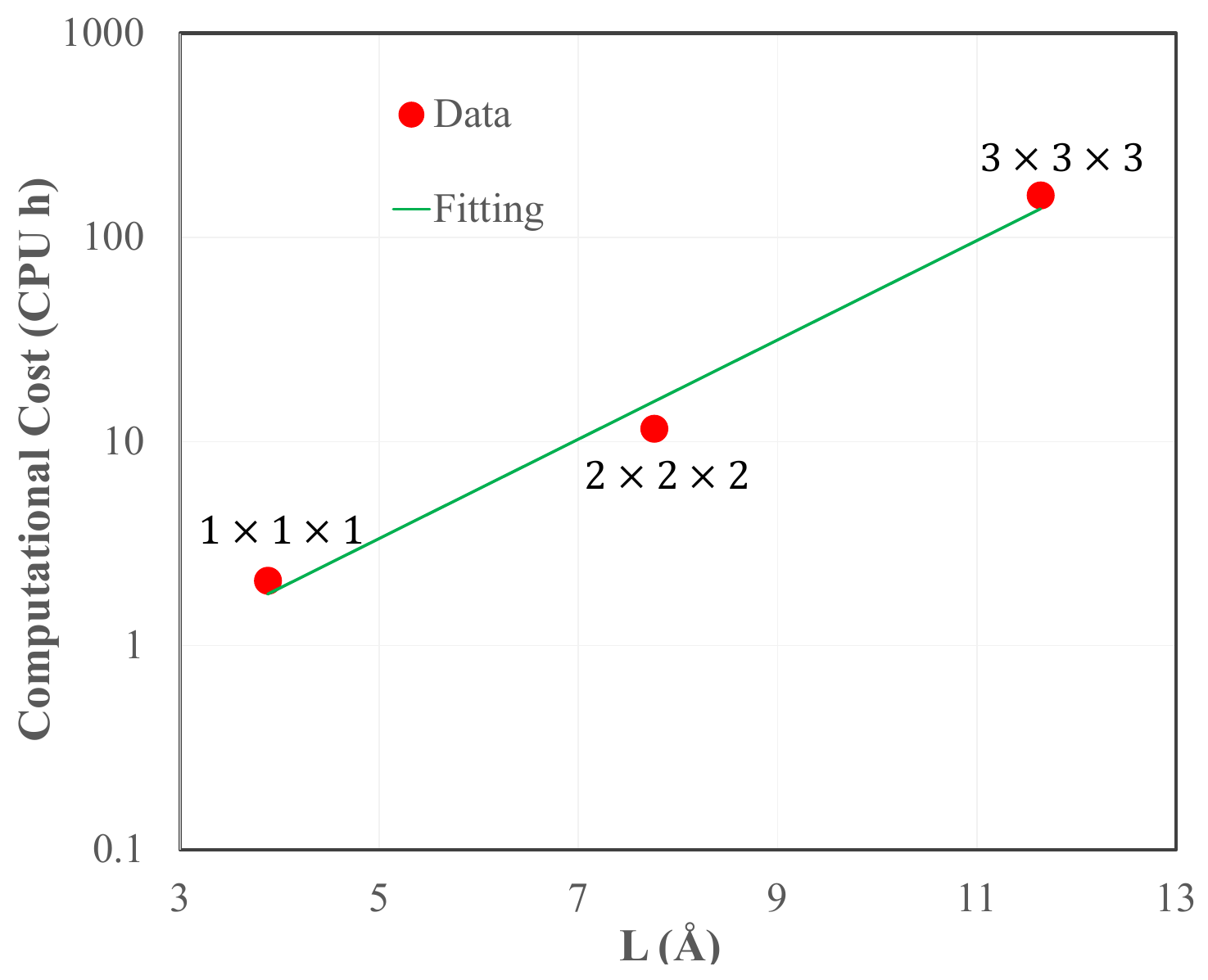}
\caption{\label{Figure_DefectiveLi3OCl_formationLivacancyComputationaltime} Dependence of the computational cost for the formation energy of lithium vacancy in Li$_{3}$OCl on the linear size of calculation supercell $L$ (Å). Each supercell is labeled next to its corresponding data point.}
\end{figure}

\noindent where $E^{(i)}$ ($i = 0, 1$) are fitting parameters. The extrapolated value of the defect formation energy for an infinite crystal, obtained by taking the limit $1/L \rightarrow 0$ in Eqn. (\ref{eqn2}), is $E_{\text{f}}(L \rightarrow \infty) = E^{(0)} = 14.39$ eV. The parameter $E^{(1)} = 1.256$ eV$\cdot$ \AA \ represents the finite-size correction of the leading-order associated with the interaction between periodically repeated vacancy images, corresponding to the term $1/L$. In contrast, the higher-order finite-size correction represented by the $ \mathcal{O}[\frac{1}{L^3}]$ term is negligible within the range of supercell sizes considered, as indicated by the excellent linear fit (\(R^2=0.997\)) of \(E_{\mathrm f}(L)\) as a function of \(1/L\). Apparently, increasing the supercell size from ($1\times1\times1$) ($L=a=3.88\text{ \AA}$) to ($2\times2\times2$) ($L=2a=7.76\text{ \AA}$) and ($3\times3\times3$) ($L=3a=11.64\text{ \AA}$) produces only marginal improvements in the accuracy of the DFT-calculated vacancy formation energy, ($E_{\mathrm{f}}(L)$), of 1.1\% (0.164 eV) and 1.5\% (0.218 eV), respectively.  

Nevertheless, as clearly seen in Figure \ref{Figure_DefectiveLi3OCl_formationLivacancyComputationaltime}, the computational cost (solid cricles), $C(L)$, increases exponentially with increasing supercell size $L$ and can be well fit ($R^{2}=0.985$) to an exponentially-growing function (solid line) of the form $C(L) = C_{0}\times10^{L/L_{0}}$, where $C_{0} = 0.2029 \text{ CPU h}$ and $L_{0} = 4.1083$ \AA. In Figure \ref{Figure_DefectiveLi3OCl_formationLivacancyComputationaltime}, the computational cost is measured in units of CPU hours (abbreviated as CPU h), a standard measure of the computational resource consumption in high-performance computing.  One CPU hour is the computational resource corresponding to one CPU core that runs continuously for one hour. The total computational cost is calculated as the product of the wall-clock runtime and the number of CPU cores employed. In order to generate each data point in Figure \ref{Figure_DefectiveLi3OCl_formationLivacancy} using Eqn. (\ref{FormationEnergyEquation}), eight CPU cores have been used for each of all DFT calculations, all of the wall-clock runtimes for each supercell size are summed and then multiplied by 8 cores to provide a data point in Figure \ref{Figure_DefectiveLi3OCl_formationLivacancyComputationaltime}.

Together, we have demonstrated that the $1\times1\times1$ supercell provides a favorable balance between computational efficiency and precision for the present defect-screening study. Although larger supercells produce slightly more accurate absolute formation energies, the improvement is marginal (<2\%), while the computational cost increases exponentially with the size of the supercell. More importantly, the $1\times1\times1$ supercell provides a consistent structural and defect environment across all dopants, enabling a systematic comparison of relative defect energetics, local structural distortions, and symmetry changes. Therefore, given the large number of substitutional defects considered in this work, we adopt the $1\times1\times1$ supercell as a computationally efficient and sufficiently accurate framework for high-throughput screening and identification of promising defect chemistries relevant to neutron-induced defect engineering.

\section{Results and Discussion \label{r&d}}

\subsection{Structural and Energetic Properties of Pristine Li$_{3}$OCl \label{R&D_Structure&Energy}}

We first optimized the pristine Li$_3$OCl unit cell using the DFT framework described above. The relaxed structure retains the cubic antiperovskite symmetry (space group $Pm\bar{3}m$) with the lattice parameter $a=b=c=3.88$\, \AA \ and the Li-O bond length 1.99 \, \AA \ (Figure \ref{Figure_Li3OCl_crystal}). These results are in good agreement with previous experimental and theoretical reports of $a =$ 3.91 \, \AA \ (Zhao and Daemen) \cite{Y_Zhao_2012} and 3.85 \, \AA \ (Zhang \textit{et al.}) \cite{Y_Zhang_2013}, respectively, and first principles calculations of the Li–O bond length 1.99 \, \AA\ \cite{A_Emly_2013,Y_Zhang_2013,M_Wu_2018}. The equilibrium cell volume is 58.58 \, \AA $^3$ and the total energy of the optimized unit cell is $-166.658$\,Ry ($-2267.1$\,eV), which we adopt as the reference energy for all calculations of defect formation.

The calculated structural parameters and total energy are consistent with previous first-principles studies \cite{A_Emly_2013,Y_Zhang_2013,M_Wu_2018}, confirming the reliability of our computational setup. These pristine-cell results provide the energetic and structural baseline used to quantify the effects of Li-site substitutional defects analyzed in the following subsections.

\subsection{Effects of Lithium-Site Substitution on Structural and Energetic Properties of Defective Li$_{3}$OCl} \label{sec3.2}}

\noindent
To assess the influence of the atomic substitution on the structural and energetic stability of Li$_3$OCl, we introduced a systematic set of substitutional species at Li lattice sites. The study includes monocations (H, Na, K), dications (Mg, Ca, Ba, He), and trication (Al, B, and Gd). For each defect, we evaluate relaxed local geometries, changes in crystallographic symmetry, and defect formation energies referenced to the pristine cell (Sect. \ref{Methods}). This comparative analysis highlights the roles of dopant charge state and effective ionic size in determining local lattice distortions, site preference, and the thermodynamic stability of Li-site substitutions in the antiperovskite framework.

\vspace{0.3cm}
\subsubsection{Structural and Bonding Modifications \label{Structural_and_Bonding}}

To systematically investigate the influence of substitutional defects on the Li$_3$OCl antiperovskite framework, we analyzed how mono-, di-, and trication dopants at Li lattice sites modify the local bonding environment and the global crystal structure. The calculated bond lengths, lattice constants, cell volumes, and symmetries are summarized in Table \ref{Table_Structure_parameter}. Overall, Li-site substitution produces both local bond rearrangement and, in many cases, symmetry lowering from the pristine cubic phase to tetragonal or orthorhombic structures. Because enhanced ionic conductivity in solid oxide electrolytes is often associated with weaker coupling between mobile Li ions and the negatively charged oxygen framework \cite{Deck_Hu_2023}, defect-induced changes in the Li--O bond length provide a useful first indication of how substitution can affect Li ion transport.

\begin{table}[H]
\centering
\caption{\label{Table_Structure_parameter} Structural and bonding modifications induced by substitutional defects at Li lattice sites in Li$_3$OCl.}
\resizebox{\textwidth}{!}{
\begin{tabular}{cccccccccccc}
\toprule
Ionic group & Dopant & $r_A$ (pm)$^{*}$ & $A$--O (\AA) & Li--O (\AA) & $a$ (\AA) & $b$ (\AA) & $c$ (\AA) & $V$ (\AA$^3$) & $\Delta V$ (\%) & Symmetry \\
\midrule
Pristine & Li & 182 & 1.99 & 1.99 & 3.88 & 3.88 & 3.88 & 58.58 & 0 & Cubic \\

\midrule
\multirow{3}{*}{Monocation}
& H & 120 & 1.94 & 2.00 & 4.92 & 5.28 & 5.48 & 142.36 & 143 & Orthorhombic \\
& Na & 227 & 2.36 & 1.93 & 4.21 & 4.21 & 4.21 & 76.62 & 27 & Cubic \\
& K & 275 & 2.73 & 2.03 & 5.29 & 5.29 & 5.29 & 148.04 & 163 & Cubic \\

\midrule
\multirow{4}{*}{Dication}
& He$^{**}$ & 140 & 1.94 & 2.00 & 2.93 & 2.93 & 4.77 & 40.95 & $-30$ & Tetragonal \\
& Ba & 268 & 2.80 & 2.13 & 5.20 & 5.20 & 5.20 & 140.61 & 140 & Cubic \\
& Ca & 231 & 2.42 & 2.02 & 5.53 & 5.53 & 5.53 & 169.11 & 189 & Cubic \\
& Mg & 173 & 2.02 & 2.14 & 3.19 & 3.19 & 5.19 & 52.81 & $-10$ & Tetragonal \\

\midrule
\multirow{3}{*}{Trication}
& B & 192 & 1.61 & 2.22 & 1.76 & 1.76 & 3.07 & 9.05 & $-85$ & Tetragonal \\
& Gd & 237 & 2.41 & 2.00 & 3.67 & 3.67 & 5.64 & 75.96 & 30 & Tetragonal \\
& Al & 184 & 1.89 & 2.27 & 4.04 & 4.04 & 4.04 & 65.45 & 12 & Cubic \\
\bottomrule \\
\end{tabular}}
\\
Note: $^{*}$ data of atomic radii (van der Waals) in picometers (pm) taken from \hyperlink{https://pubchem.ncbi.nlm.nih.gov/periodic-table/atomic-radius/}{National Center for Biotechnology Information}.\\
$^{**}$ Helium ion is a dication ion, i.e., an alpha particle.
\end{table}

Among the monocation defects, Na and K preserve the cubic symmetry of Li$_3$OCl, but both expand the lattice substantially. The large ionic radius of K results in the strongest isotropic expansion in this group, increasing the unit-cell volume by 163\%. It also increases the K--O bond length to 2.73 \AA\ and slightly elongates the neighboring Li--O bond to 2.03 \AA. This Li--O bond elongation suggests weaker Coulomb coupling between Li ions and the oxygen framework, which may promote Li-ion migration by softening the lithium sublattice. In contrast, Na substitution shortens the Li--O bond to 1.93 \AA, implying somewhat stronger Li--framework coupling despite the expanded cell.

Hydrogen behaves very differently from the larger alkali-metal dopants. Although H is also monocation, its small ionic radius induces pronounced anisotropic distortion and lowers the symmetry from cubic to orthorhombic. The resulting lattice constants become highly unequal ($a=4.92$ \AA, $b=5.28$ \AA, and $c=5.48$ \AA), and the unit-cell volume increases by 143\%. At the same time, the H--O bond contracts to 1.94 \AA, indicating strong localized H--O bonding, whereas the neighboring Li--O bond expands slightly to 2.00 \AA. This combination points to local bonding reconstruction around H together with partial decoupling of nearby Li from the oxygen framework. Since H is a transmutation product associated with neutron irradiation of lithium-containing materials, these results suggest that neutron-generated hydrogen or tritium incorporation may strongly perturb the local bonding topology and structural stability of Li$_3$OCl. 

The dication defect set shows a stronger competition among ionic size, lattice strain, and local bonding. Large alkaline-earth dopants such as Ba and Ca preserve cubic symmetry, but induce very large isotropic expansion. In particular, Ca produces the largest unit-cell volume of all defects considered, corresponding to a 189\% increase relative to pristine Li$_3$OCl.  The corresponding Li--O bond lengths of 2.02--2.13 \AA\ indicate appreciable weakening of Li--framework interactions, which could improve Li-ion transport by increasing lattice softness and widening migration pathways.

In contrast, He and Mg both relax to tetragonal structures. Helium is especially unusual because it contracts the lattice in the $ab$ plane while elongating the $c$ axis, yielding an overall volume reduction of 30\%. Although the He--O distance is short (1.94 \AA), He is chemically inert and cannot form stabilizing directional bonds. The structural response is therefore best understood as anisotropic lattice collapse rather than conventional bonding. Since the neighboring Li--O bonds remain essentially unchanged at 2.00 \AA, He appears to introduce structural disorder more than effective Li--O decoupling. Since He is also a transmutation product generated under neutron irradiation of Li- and B-containing systems, its incorporation may contribute to local symmetry breaking and radiation-induced disorder in Li$_3$OCl.

Mg likewise induces tetragonal distortion, but with only a moderate volume contraction of 10\%. More importantly, Mg gives one of the longest Li--O bonds in the entire defect set (2.14 \AA). This pronounced elongation indicates significant weakening of Li--framework coupling and suggests that Mg substitution may be particularly effective in modifying the local environment relevant to Li-ion migration, despite the modestly reduced lattice volume.

The trication defects exhibit the strongest competition between local bonding reconstruction and electrostatic mismatch. Al preserves cubic symmetry and causes only moderate volume expansion (12\%), yet it produces the largest Li--O bond elongation of all cases examined, reaching 2.27 \AA. This result points to a substantial weakening of Li--framework interactions and suggests that Al substitution may strongly influence the migration landscape for Li ions.

Boron produces the most extreme structural reconstruction. The lattice collapses into a highly compressed tetragonal structure with an 85\% reduction in unit-cell volume, while the B--O bond shortens dramatically to 1.61 \AA. 
In contrast, the neighboring Li--O bond expands to 2.22 \AA. These changes indicate that B strongly localizes bonding at the defect center while simultaneously weakening Li--O interactions in the surrounding lattice. Since B is directly relevant to neutron-absorbing environments, this result suggests that boron incorporation can fundamentally reconstruct the local bonding topology of Li$_3$OCl even before neutron irradiation. The strong B--O bond may act as a rigid anchoring center, whereas the elongated Li--O bonds nearby may favor enhanced Li mobility in adjacent regions.

Gd also lowers the symmetry to tetragonal, but its effect is less localized than that of B. Rather than driving strong bond reconstruction at the defect center, the large ionic radius and high charge of Gd primarily generate long-range anisotropic strain. As a result, the Li--O bond remains close to the pristine value, while the lattice expands markedly along the $c$ direction.

Taken together, these results show that substitutional defect engineering in Li$_3$OCl is governed by a delicate interplay of ionic size, charge state, local bonding chemistry, and symmetry-lowering lattice relaxation. Several defects, especially K, Ba, Ca, Mg, Al, and B, significantly elongate the Li--O bond relative to pristine Li$_3$OCl, indicating weakened Li--framework coupling. Such decoupling may be a useful structural descriptor for fast Li-ion transport in oxide solid electrolytes. However, its predictive value must still be tested against explicit transport calculations. In particular, nudged elastic band (NEB) calculations and large-scale molecular dynamics simulations are needed to determine whether Li--O bond elongation alone can serve as a reliable descriptor of lithium transport, or whether it must be considered together with migration barriers along minimum-energy pathways that govern the Li diffusion coefficient. Overall, the present results suggest that neutron-transmutation-related defects, particularly H and B, may offer unique non-equilibrium routes for tuning local bonding topology, lattice softness, and potentially Li-ion mobility in antiperovskite solid electrolytes.

Beyond the structural distortions and bonding modifications discussed in Section \ref{Structural_and_Bonding}, the energetic stability of the substitutional defects provides additional insight into the thermodynamic response of Li$_3$OCl to  doping and neutron-transmutation-related species. In particular, the defect formation energetics complement the previously identified trends in lattice expansion, symmetry lowering, and Li--O bond modification. Defects that strongly elongate the Li--O bond and induce substantial lattice relaxation, such as H, Mg, Al, and B may locally weaken Li--framework coupling and create structurally favorable environments for Li transport, but their practical relevance ultimately depends on whether these configurations are thermodynamically accessible and kinetically persistent in the antiperovskite lattice. Highly unfavorable formation energies would imply that some of these defects are more likely to occur as transient non-equilibrium species under neutron irradiation than as stable equilibrium dopants.

\subsubsection{Substitutional Defect Formation Energetics \label{formationEnergy}}

\noindent
Beyond the structural distortions discussed above, the energetic stability of the substitutional defects provides further insight into the thermodynamic response of Li$_3$OCl to substitutional defects and neutron-transmutation-related species. Table \ref{tab:substitutional_defects} summarizes the calculated defect formation energies for all substitutional configurations grouped according to dopant valence.

\begin{table}[htbp]
\centering
\caption{Substitutional defect formation energies in Li$_3$OCl.}
\label{tab:substitutional_defects}
\renewcommand{\arraystretch}{1.3}
\begin{tabular}{ccc}
\toprule
\textbf{Valency} & \textbf{Dopant} & \textbf{Formation Energy (eV)} \\
\midrule

\multirow{3}{*}{Monocation}
& H  & 20.00 \\
& Na & 8.69 \\
& K  & 8.14 \\
\midrule

\multirow{4}{*}{Dication}
& He & 4.54 \\
& Ba & 7.81 \\
& Ca & 8.09 \\
& Mg & 7.18 \\
\midrule

\multirow{3}{*}{Trication}
& B  & 4.57 \\
& Gd & 7.44 \\
& Al & 6.21 \\
\bottomrule

\end{tabular}
\end{table}

\textbf{Monocation substitutional defects}. Among the monocation substitutional species, hydrogen exhibits an exceptionally large magnitude of defect stabilization energy (20 eV) within the present reference framework based on molecular hydrogen, corresponding to approximately 16 eV per H atom (derived from an H$_2$ total energy of $\sim$32 eV per molecule). This value is substantially larger in magnitude than that of all other dopants considered in this study, indicating a strong energetic driving force for hydrogen incorporation at the Li site within the adopted substitutional formalism. In contrast, use of a hypothetical crystalline hydrogen reference state yields a substantially larger magnitude (on the order of 64 eV), which is not physically meaningful under standard thermodynamic conventions and highlights the strong sensitivity of the calculated energy to the choice of reference state. Overall, these results indicate a strong thermodynamic driving force for H incorporation at Li lattice sites within the imposed computational framework. Such behavior likely originates from the very small effective ionic radius of hydrogen, its strong local electrostatic interaction with surrounding oxygen atoms, and the ability of the lattice to undergo substantial local relaxation to accommodate H-related bonding configurations.

However, hydrogen incorporation in Li$_3$OCl is not strictly equivalent to a simple substitutional defect in a rigid cubic framework. This is understandable because, in real systems sufficiently high H concentrations may drive local structural reconstruction involving O--H bond formation, and may favor paddle-wheel dynamics and potential transformation toward hydroxylated phases such as Li$_2$OHCl, which is known to adopt crystal structures distinct from the cubic antiperovskite phase. Therefore, the computed energetic stabilization should be interpreted primarily as an indicator of a strong thermodynamic driving force for H incorporation rather than as a definitive prediction of phase stability within the Li$_3$OCl lattice.

In contrast, the alkali-metal dopants Na and K exhibit substantially smaller stabilization magnitudes (8.69 eV and 8.14 eV, respectively), consistent with more conventional substitutional behavior dominated primarily by ionic-size mismatch and lattice-strain effects, with minimal tendency toward reconstructive chemistry.

From a radiation perspective, these results gain additional significance under neutron irradiation conditions. In particular, neutron transmutation of $^{6}$Li produces energetic tritons and alpha particles, generating localized MeV-scale collision cascades that can create transient thermal spikes and highly non-equilibrium defect environments. Within such cascade cores, the lattice may undergo momentary amorphization or even localized melting on sub-picosecond timescales, followed by rapid quenching. Under these conditions, hydrogen- or tritium-rich local environments may become kinetically trapped in configurations that are not accessible under equilibrium thermodynamic synthesis pathways.

At room-temperature irradiation conditions, however, the relaxation dynamics are expected to favor the formation of highly disordered or defect-rich amorphous-like regions rather than fully ordered crystalline hydroxylated phases. Consequently, while hydroxylated or H-rich phases such as Li$_2$OHCl may represent thermodynamically relevant endpoints under equilibrium conditions, neutron irradiation in insulating Li$_3$OCl is more likely to produce kinetically frozen defect configurations and locally disordered regions governed by cascade-driven non-equilibrium processes rather than complete structural recrystallization.

\textbf{Dication substitutional defects}. The dication substitutional species generally exhibit less favorable energetics than the most strongly stabilized monocation case, reflecting the increasing electrostatic mismatch and stronger local charge perturbation introduced by aliovalent substitution at the Li site. Among the conventional dication dopants, Ca and Ba exhibit relatively large stabilization magnitudes of 8.09 eV and 7.81 eV, respectively, suggesting that significant lattice expansion and local structural relaxation can partially offset the electrostatic penalty associated with dication substitution in the Li$_3$OCl antiperovskite framework. Mg also exhibits substantial stabilization (7.18 eV) according to the calculated defect energetics summarized in Table \ref{tab:substitutional_defects}, although its exact relative ordering should be interpreted in conjunction with the adopted chemical reference state.

In contrast, He exhibits one of the least favorable formation energies among all dopants investigated (4.54 eV), indicating weak energetic compatibility with the Li$_3$OCl lattice. This behavior is consistent with the chemically inert nature of helium and its inability to form stabilizing chemical interactions or participate in directional bonding within the host lattice.

Importantly, helium is not merely a hypothetical dopant species in this context, but directly relevant to neutron-induced transmutation processes. In particular, thermal neutron capture by $^{6}$Li and $^{10}$B isotopes generates highly energetic nuclear reaction products, including MeV-scale $\alpha$-particles ($^{4}$He nuclei) and tritons ($^{3}$H). These transmutation events produce localized helium incorporation and displacement cascades within the lattice, thereby creating non-equilibrium defect environments far beyond conventional thermodynamic doping regimes. The strongly unfavorable substitutional energetics obtained here for He therefore reflect its intrinsic inability to stabilize within the Li$_3$OCl framework, while also highlighting its role as a transient but physically significant species in neutron-irradiated solid electrolytes. Such He-related defect states are expected to contribute primarily to radiation-induced lattice disorder, defect clustering, and cascade-driven structural evolution rather than equilibrium substitutional stabilization.

\textbf{Trication substitutional defects}. The trication dopants exhibit intermediate energetic stability governed by a coupled interplay between ionic radius, charge imbalance, and local bonding reconstruction within the Li$_3$OCl antiperovskite lattice. Aluminum and gadolinium yield formation energies of 6.21 eV and 7.44 eV, respectively, indicating moderate energetic stability despite the substantial electrostatic perturbation associated with substitution at the Li sites. In particular, Gd remains energetically more favorable than He and comparable to other strongly mismatched species, despite inducing significant local structural distortion. This behavior suggests that the Li$_3$OCl lattice can partially accommodate high-valence dopants through volumetric relaxation and oxygen-sublattice rearrangement.

Boron, despite its strong bonding propensity, exhibits a less favorable formation energy of 4.57 eV. This indicates that local bonding stabilization alone is insufficient to compensate for the strong electrostatic penalty associated with trication substitution at the Li sites, and highlights the dominant role of charge imbalance in governing defect energetics within this framework.

As discussed in the Introduction, beyond purely chemical substitution, both B and Gd are of particular relevance in the context of neutron-induced defect engineering due to their distinct nuclear interaction characteristics. Natural lithium contains only 7.5\% of the $^{6}$Li isotope, transmuted by neutrons via $^{6}\mathrm{Li}+n\rightarrow\,^{3}\mathrm{H}+\alpha$, which has a thermal neutron absorption cross section of approximately 940 barns. In contrast, upon neutron capture, $^{10}$B undergoes the $^{10}\mathrm{B}+n\rightarrow\,^{7}\mathrm{Li}+\alpha+\gamma$ reaction, producing energetic $\alpha$-particles ($^{4}$He$^{2+}$), recoiling $^{7}$Li ions, and MeV-energy $\gamma$-photons. Naturally abundant $^{10}$B (20\% abundance) exhibits a much larger thermal neutron capture cross section of 3,840 barns, making boron a far more efficient neutron absorber in mixed ionic solids. Thus, even a small concentration of boron dopants in Li$_3$OCl can act as highly efficient, spatially localized neutron absorption centers, enabling targeted defect generation without requiring isotopic enrichment of lithium, a process that is both costly and challenging.

In contrast, gadolinium interacts with neutrons primarily through radiative capture reactions, dominated by the isotopes $^{157}$Gd and $^{155}$Gd, which possess exceptionally large thermal neutron capture cross sections of approximately $2.55 \times 10^{5}$ barns and $6.0 \times 10^{4}$ barns, respectively. Following neutron absorption, the excited compound nucleus relaxes via emission of prompt gamma rays, internal conversion electrons, and characteristic X-rays. In insulating polycrystalline materials, such energetic electronic excitations can lead to local ionization, space-charge redistribution, and grain-boundary charge neutralization. These processes have been shown to reduce electrostatic barriers at grain boundaries, thereby lowering Li migration barriers and enhancing overall ionic conductivity through improved intergranular transport pathways \cite{Thomas_Defferriere_2024,Thomas_Defferriere_2025}.

Taken together, these nuclear and electronic interaction pathways provide the physical basis for selecting B and Gd as representative trication dopants in this study. Boron enables localized MeV-scale displacement cascade formation via $(n,\alpha)$ reactions, while gadolinium enables gamma-mediated electronic excitation and grain boundary charge modulation. These fundamentally distinct neutron–matter interaction mechanisms motivate their inclusion in the present defect energetics framework, providing a direct link between substitutional chemistry, neutron physics, and defect-engineered ionic transport in Li$_3$OCl.

\section{Conclusions and Outlook}

In this work, we performed a systematic first-principles investigation of substitutional lattice defects in Li$_{3}$OCl associated with neutron-transmutation-related species and conventional dopants. The DFT calculations reveal that substitutional defects strongly modify the structural stability, local bonding environment, lattice symmetry, and defect energetics of the antiperovskite framework. In particular, neutron-related species such as H, He, and B induce pronounced local structural distortions and symmetry lowering, demonstrating that neutron transmutation can fundamentally alter the defect chemistry of Li$_{3}$OCl beyond equilibrium doping regimes.

The results further show that substitutional defects can significantly modify Li--O bonding environments, which may influence Li-ion transport by weakening Li--framework coupling and altering lattice softness. Several defects, including H, Mg, Al, and B, produce substantial Li--O bond elongation and strong local lattice relaxation, suggesting that neutron-induced defects may provide non-equilibrium pathways for tuning ionic transport in antiperovskite solid electrolytes. At the same time, the calculated defect formation energetics indicate that the stability of these defects is governed by a complex interplay among ionic radius, charge imbalance, local bonding reconstruction, and lattice strain. In particular, He-related defects are energetically unfavorable, supporting the interpretation that helium generated during neutron irradiation primarily contributes to transient lattice disorder, defect clustering, and radiation-induced structural instability rather than equilibrium substitutional stabilization.

Beyond defect chemistry, this study establishes a direct conceptual link between neutron physics and defect-engineered ionic transport in Li$_{3}$OCl. Defects associated with neutron capture by $^{10}$B provide a model system for localized displacement cascade generation, while H and He represent transmutation products generated from neutron capture by $^{6}$Li. These neutron-induced species may therefore create highly non-equilibrium defect environments inaccessible through conventional synthesis routes, providing new opportunities for irradiation-assisted materials engineering in solid-state battery materials.

Several limitations of the present study should be acknowledged. The calculations were performed using a reduced $1\times1\times1$ supercell and a simplified defect formation formalism without explicit charged-defect corrections, finite-temperature contributions, or dilute-limit extrapolation. Consequently, the calculated formation energies should primarily be interpreted as comparative descriptors of relative defect stability and local structural response rather than as rigorous equilibrium thermodynamic quantities.

Future work will extend this study using larger supercells, nudged elastic band (NEB) calculations, \textit{ab initio} molecular dynamics (AIMD), and electronic band structure calculations to directly investigate lithium-ion and electron transport in Li$_{3}$OCl. More broadly, the present work demonstrates that neutron-driven defect engineering may provide a promising strategy for designing advanced solid electrolytes and other energy materials for next-generation all-solid-state lithium-ion batteries.

\section*{Conflicts of Interest} 
The authors declare no conflict of interest.

\section*{Author Contributions}

The contributions of the author to the current work are as follows:

Conceptualization: H. M. N., C.W., T. W. H, Y. X, and J. G.

Methodology: H. M. N., C. D. Z., C. W., N.N., A. D., and C. G.

Manuscript Writing:  C. D. Z., H.M.N., C.W.

Manuscript Proofreading and Reviewing: all authors.

Equal Contributions: H. M. N and C. D.Z.

\section*{Acknowledgments}
  This work was funded in part by the National Science Foundation Grant No. IIP-2044726, the University of Missouri Materials Science and Engineering Institute (MUMSEI) Grant No. CD002339, and the University of Missouri Research Reactor (MURR). The computation for this work was performed on the high-performance computing infrastructure provided by Research Support Services at the University of Missouri, Columbia, MO. DOI: https://doi.org/10.32469/10355/97710. 



\end{document}